\def \SAIT #1 #2 {{\em Mem.\ Soc.\ Astron.\ It.\/} {\bf #1}, #2}
\def \MESS #1 #2 {{\em The Messenger\/} {\bf #1}, #2}
\def \ASTRNACH #1 #2 {{\em Astron. Nach.\/} {\bf #1}, #2}
\def \AAP #1 #2 {{\em Astron. Astrophys.\/} {\bf #1}, #2}
\def \AAL #1 #2 {{\em Astron. Astrophys. Lett.\/} {\bf #1}, L#2}
\def \AAR #1 #2 {{\em Astron. Astrophys. Rev.\/} {\bf #1}, #2}
\def \AAS #1 #2 {{\em Astron. Astrophys. Suppl. Ser.\/} {\bf #1}, #2}
\def \AJ #1 #2 {{\em Astron. J.\/} {\bf #1}, #2}
\def \ANNREV #1 #2 {{\em Ann. Rev. Astron. Astrophys.\/} {\bf #1}, #2}
\def \APJ #1 #2 {{\em Astrophys. J.\/} {\bf #1}, #2}
\def \APJL #1 #2 {{\em Astrophys. J. Lett.\/} {\bf #1}, L#2}
\def \APJS #1 #2 {{\em Astrophys. J. Suppl.\/} {\bf #1}, #2}
\def \APSS #1 #2 {{\em Astrophys. Space Sci.\/} {\bf #1}, #2}
\def \ASR #1 #2 {{\em Adv. Space Res.\/} {\bf #1}, #2}
\def \BAIC #1 #2 {{\em Bull. Astron. Inst. Czechosl.\/} {\bf #1}, #2}
\def \JSQRT #1 #2 {{\em J. Quant. Spectrosc. Radiat. Transfer\/} {\bf #1}, #2}
\def \MN #1 #2 {{\em Mon. Not. R. Astr. Soc.\/} {\bf #1}, #2}
\def \MEM #1 #2 {{\em Mem. R. Astr. Soc.\/} {\bf #1}, #2}
\def \PLR #1 #2 {{\em Phys. Lett. Rev.\/} {\bf #1}, #2}
\def \PASJ #1 #2 {{\em Publ. Astron. Soc. Japan\/} {\bf #1}, #2}
\def \PASP #1 #2 {{\em Publ. Astr. Soc. Pacific\/} {\bf #1}, #2}
\def \NAT #1 #2 {{\em Nature\/} {\bf #1}, #2}

\documentstyle{memsait}
\input epsf.sty
\begin{opening}
\title{AN AUTOMATIC PROCEDURE TO EXTRACT GALAXY CLUSTERS 
FROM CRONARIO CATALOGUES}
\author{E. PUDDU$^1$, S. ANDREON$^1$, G. LONGO$^1$, M. PAOLILLO$^2$,
R. SCARAMELLA$^3$, V. TESTA$^3$, R.R. GAL$^4$, R.R. DE CARVALHO$^5$,
S.G. DJORGOVSKI$^4$}
\institute{$^1$Osservatorio Astronomico di Capodimonte, Napoli, Italy\\
$^2$Osservatorio Astronomico di Palermo, Palermo, Italy\\
$^3$Osservatorio Astronomico di Monte Porzio, Roma, Italy\\
$^4$Palomar Observatory, Caltech, Pasadena, CA\\
$^5$Observatorio Nacional/CNPq, Brazil}
\date{} 
\end{opening}

\begin{document}

\oddpagefooter{}{}{} 
\evenpagefooter{}{}{} 
\ 
\bigskip

\begin{abstract}
We present 
preliminary results of a simple peak finding algorithm applied to
catalogues of galaxies, extracted from the Second Palomar Sky Survey
in the framework of the CRoNaRio project. 
All previously known Abell and Zwicky clusters in a test region of
$5^{\circ} \times 5^{\circ}$ are recovered 
and new candidate clusters are also detected.\\ 
This algorithm represents an alternative way of searching for galaxy
clusters  with respect to that implemented and tested at Caltech on
the same type of data (Gal et al. 1998).
\end{abstract}

\section{Introduction}
Clusters of galaxies are the largest known virialized structures and 
represent the high-density peaks of the large scale structure of the
Universe which is effectively traced up to 150h$^{-1}$ Mpc or more (Bahcall \&
Soneira 1984; Tully 1987). 
The distribution and the evolution of intrinsic properties of galaxy
clusters provide important information for studies of galaxy and 
cluster evolution and on the dependence of galaxy properties on
the environment.\\
In the past, a number of cluster catalogues (cf. Abell 1958; Abell et al. 
1989; Zwicky et al. 1961-68) has been extracted
from photographic all-sky surveys. These catalogues, however, were compiled by 
visual inspection of the plates and lack the homogeneity and 
completeness which is needed for statistical studies (Postman et al. 1986; 
Sutherland 1988).\\ 
Machine extracted catalogues (Dodd \& MacGillivray 1986; Dalton et al. 1992;
Lumsden et al. 1992) selected with more objective criteria, 
reach, in some cases, fainter limiting magnitudes but do not cover
equally wide areas of the sky.
More recently, CCD surveys have been carried out
(Couch et al. 1991; Postman et al. 1996; Olsen et al. 1999), but they
cover small regions of the sky and target deeper objects.\\

\section{The CRoNaRio data}
The CRoNaRio project (Djorgovski et al. 1998) is a joint enterprise between 
Caltech and the astronomical observatories of Monte Porzio, Napoli and Rio de 
Janeiro, aimed to the production of the Palomar Norris Sky Catalogue 
(Djorgovski et al. 1999), 
which will eventually include all objects visible on the Second Palomar Sky Survey
and therefore will provide a large database from which to extract
a statistically well defined catalogue of putative galaxy clusters (see
Gal et al. 1999).
For each POSS-II field the CRoNaRio project provides astrometric, geometric
and photometric information (J, F and N bands, calibrated through CCD
frames in the g, r, i bands of the Gunn-Thuan system).

The procedure for the search of cluster candidates, discussed in this
paper, has been developed at the
Astronomical Observatory of Capodimonte (Naples) and differs in many points 
from the standard way of preparing and generating the DPOSS distributable 
catalogues.
The first step of our procedure requires the cleaning of the catalogues
from spurious objects and artifacts (such as multiple detections coming from 
extended patchy objects, halos of bright stars, satellite tracks, etc.), 
which are present in the single filter catalogues.
We mask and keep memory of the plate regions occupied by bright, extended
and saturated objects (that locally make the detection extremely unreliable),
taking this troubled areas into account in the next steps.
Subsequently, the procedure performs the matching of the catalogues 
of the same sky region obtained in the three bands. 
This is done through the plate astrometric solution, matching each object in one filter 
with the nearest objects in the two other filters and assuming a tolerance box of 7 arcsec.

\begin{figure}
\hspace{0.5cm}
\epsfysize=4.cm
\epsfxsize=4.cm
\epsfbox{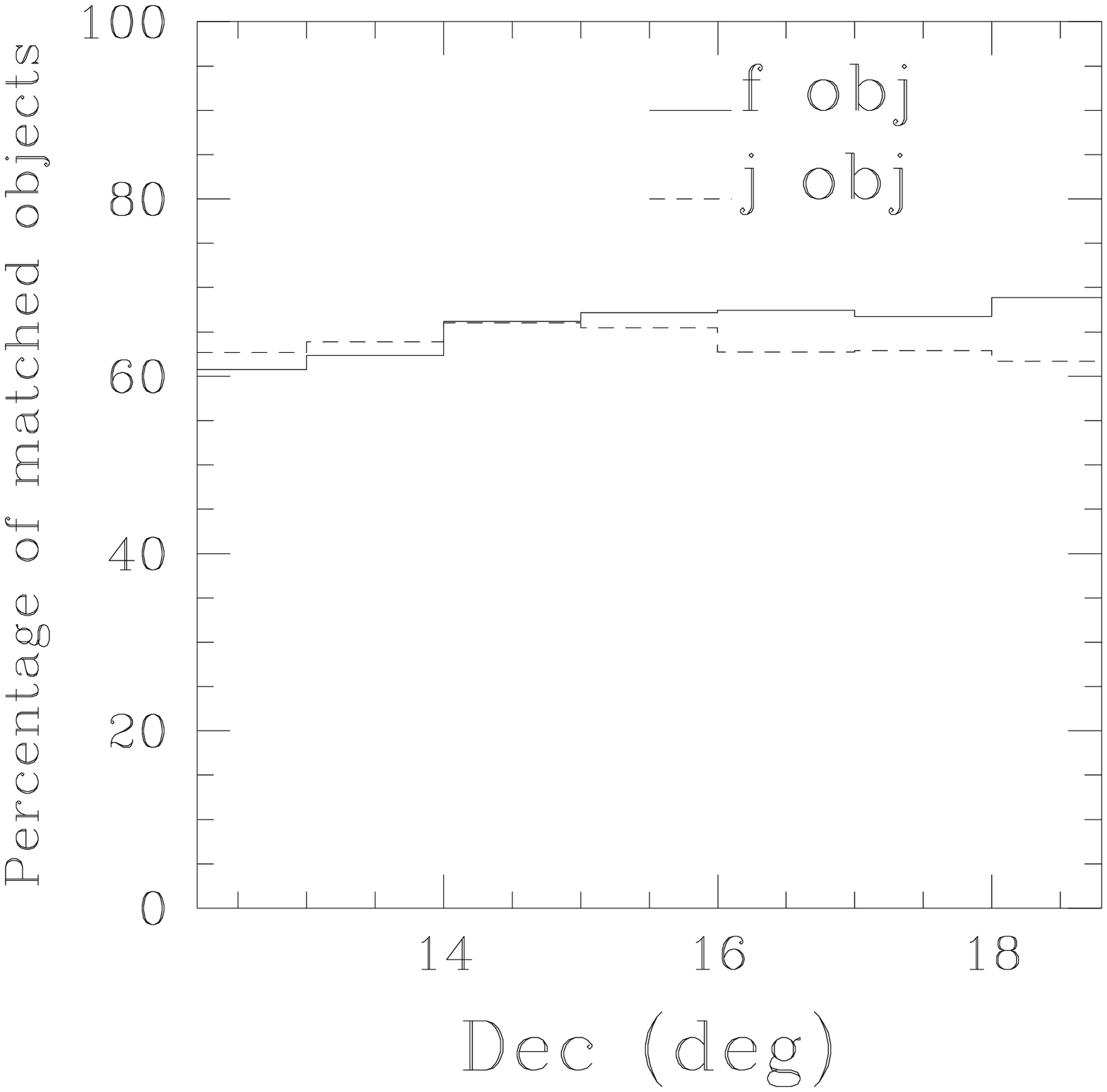}  
\epsfysize=4.cm
\epsfxsize=4.cm
\epsfbox{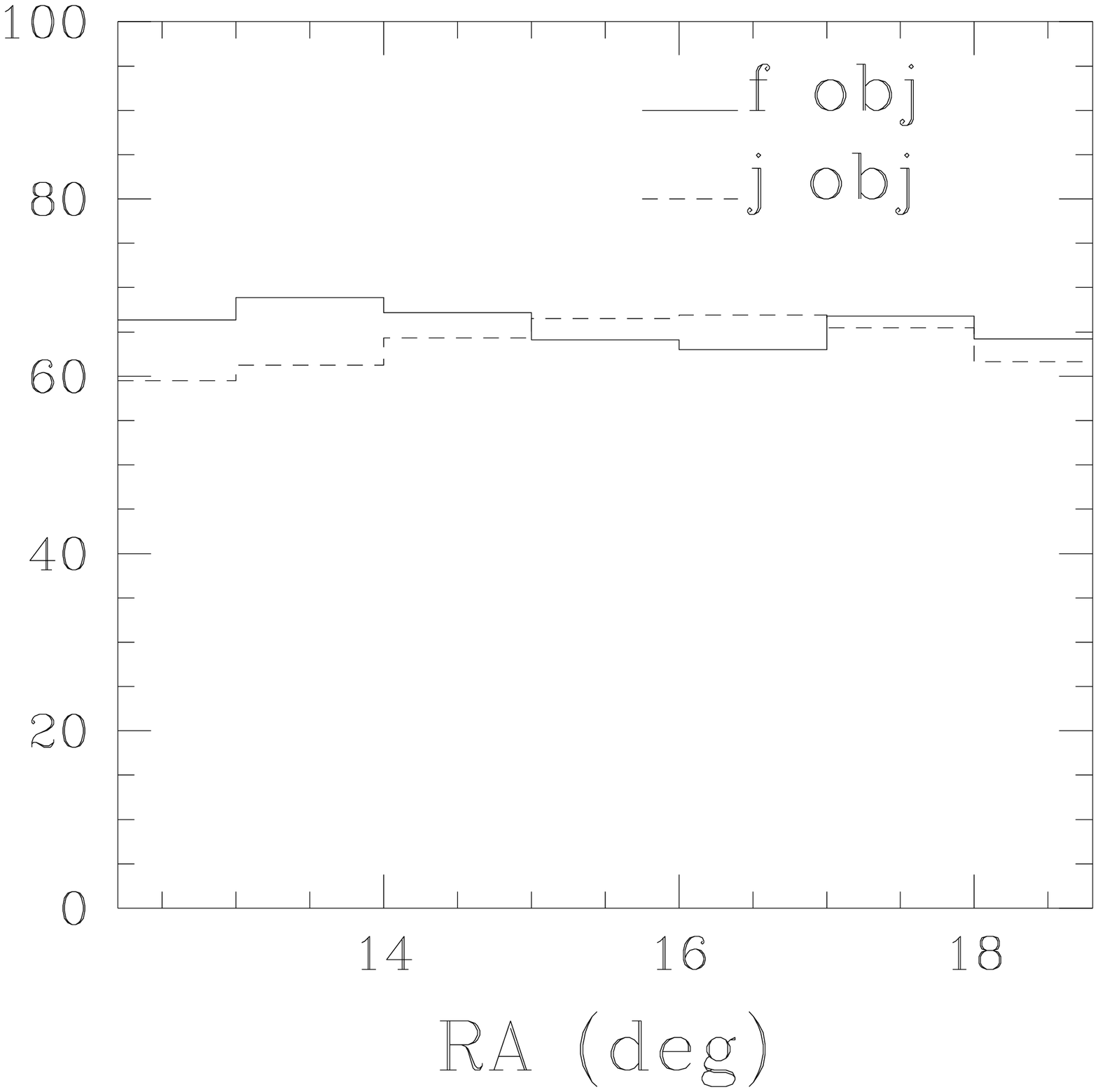} 
\epsfysize=4.cm
\epsfxsize=4.cm
\epsfbox{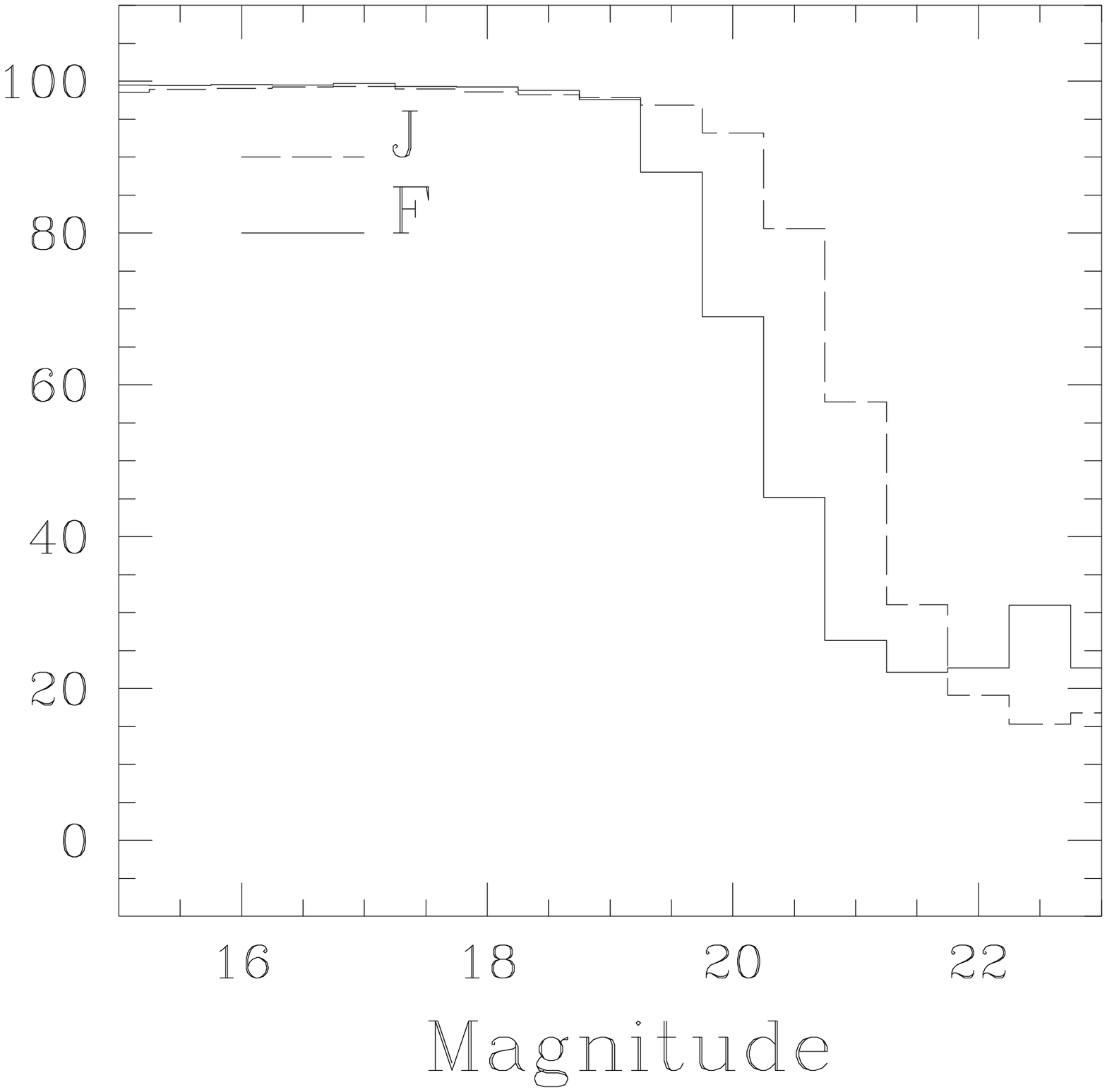} 
\caption[h]{Fraction of matched F-J objects versus the original F (continue 
line) and J (dashed line) objects as a function of right ascension (left), 
declination (center), and J and F instrumental magnitudes (right).}
\end{figure}

The quality of the matching does not depend on the position of the objects on the plates:
the fraction of the matched objects with respect to the original single filter catalogue
is quite constant all over the plate (Fig.1, left and central panel).
The quality of the matching depends, as expected, on the instrumental magnitude: 
at faint magnitude a significative fraction of objects (too faint to be detected in some
among the three filter) is lost in the matching process (Fig.1, right panel).\\
Due to the different S/N ratios in the three bands, many objects have discordant 
star-galaxy classifications in catalogues from different filters. The number of such 
objects obviously increases at faint magnitudes. 
(It needs to be stressed, however, that this problem 
is greatly reduced when a new training set for the classification is
adopted, see Gal et al. 1999 for details).

\begin{figure}[h]
\epsfysize=9cm 
\hspace{1.5cm}\epsfbox{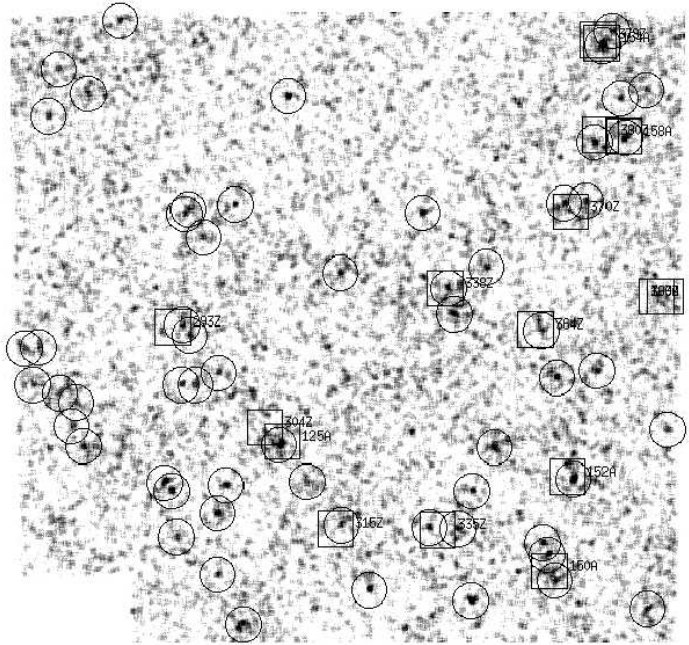} 
\caption[h]{Detected overdensities in the smoothed map in the plate n. 610. 
Squares mark known Abell and Zwicky clusters. Circles mark putative - previously
unknown - clusters.}
\end{figure}

\section{The identification of cluster candidates}
In what follows we shall refer to POSS-II field n. 610 ($5^{\circ} \times
5^{\circ}$ centered at RA = 1h and $\delta$ =$+15^{\circ}$).
After matching and taking into account the above mentioned problem
of misclassified objects, we first produced a catalogue of galaxies 
which is almost complete down to F $\sim$ 19.75 and J $\sim$ 20.21 mag.   
The spatial galaxy distribution in this catalogue was then binned into equal-area 
square bins of 36 arcsec in the sky and the resulting density map
was smoothed with a Gaussian 2-D filter having a variable 
width chosen in function of the estimated cluster redshift in
order to have a $\sim$ 250 kpc resolution (i.e., the dimension of the core 
radius of a typical cluster). 
Then, SExtractor (Bertin \& Arnouts 1996) was run on the smoothed map in order 
to identify and extract all overdensities having number density $3\sigma$ above the mean 
background and covering at least 16 pixels (9.6 arcmin).
In this way, as in the Schectman (1985) approach, we are  not 
assuming any {\it a priori} cluster model.\\
All the previously known Abell, Zwicky or X-ray clusters 
were recovered and many new candidates were detected (see Fig.2). 

\section{Conclusions}
We implemented a simple, but well understood and model independent,
procedure for cleaning CRoNaRio catalogues from spurious sources.
The procedure has been used to select the galaxy catalogues used for the determination 
of the LF of galaxy clusters and for the search of poor galaxy groups (Paolillo et al.; 
De Filippis et. al.,these Proceedings).\\
Our next goal is to validate our detections by cross-identification with
X ray catalogues of galaxy clusters or by direct optical observation.


\end{document}